\documentstyle[12pt]{article} 
\newcommand{\leqsim}{\,\raisebox{-0.6ex}{$\buildrel < \over \sim$}\,}
\newcommand{\geqsim}{\,\raisebox{-0.6ex}{$\buildrel > \over \sim$}\,}
\def\c{{\rm c}}
\def\dec{{\rm dec}}
\def\deS{{\rm deS}}
\def\df{{\rm d}}
\def\e{{\rm e}}
\def\eq{{\rm eq}}
\def\en{{\rm end}}
\def\etal{{\em et al}\,}
\def\for{{\rm for}}
\def\H{{\rm H}}
\def\I{{\rm I}}

\def\in{{\rm ini}}
\def\O{{\cal O}}
\def\P{{\rm Pl}}
\def\R{{\rm R}}
\def\rms{{\rm rms-PS}}
\def\s{{\rm s}}
\def\tot{{\rm tot}}
\def\ap#1#2#3   {{\em Ann. Phys.} {\bf#1} (#2) #3}
\def\apj#1#2#3  {{\em Astrophys. J.} {\bf#1} (#2) #3}
\def\app#1#2#3  {{\em Acta. Phys. Pol.} {\bf#1} (#2) #3}
\def\araa#1#2#3 {{\em Ann. Rev. Astron. Astrophys.} {\bf#1} (#2) #3}
\def\arnp#1#2#3 {{\em Ann. Rev. Nucl. Part. Sci.} {\bf#1} (#2) #3}
\def\cmp#1#2#3  {{\em Comm. Part. Phys.} {\bf#1} (#2) #3}
\def\err#1#2#3  {{\em Erratum} {\bf#1} (#2) #3}
\def\ib#1#2#3   {{\em ibid} {\bf#1} (#2) #3}
\def\jmp#1#2#3  {{\em J. Math. Phys.} {\bf#1} (#2) #3}
\def\ijmp#1#2#3 {{\em Int. J. Mod. Phys} {\bf#1} (#2) #3}
\def\jetp#1#2#3 {{\em JETP Lett.} {\bf#1} (#2) #3}
\def\jpg#1#2#3  {{\em J. Phys. G.} {\bf#1} (#2) #3}
\def\mn#1#2#3   {{\em Mon.~Not.~R.~astr.~Soc.} {\bf#1} (#2) #3}
\def\mpl#1#2#3  {{\em Mod. Phys. Lett.} {\bf#1} (#2) #3}
\def\nat#1#2#3  {{\em Nature} {\bf#1} (#2) #3.}
\def\nc#1#2#3   {{\em Nuovo Cim.} {\bf#1} (#2) #3}
\def\nim#1#2#3  {{\em Nucl. Instr. Meth.} {\bf#1} (#2) #3}
\def\np#1#2#3   {{\em Nucl. Phys.} {\bf#1} (#2) #3}
\def\pcps#1#2#3 {{\em Proc. Cam. Phil. Soc.} {\bf#1} (#2) #3}
\def\pl#1#2#3   {{\em Phys. Lett.} {\bf#1} (#2) #3}
\def\pr#1#2#3 	{{\em Phys. Rev.} {\bf#1} (#2) #3}
\def\prep#1#2#3 {{\em Phys. Rep.} {\bf#1} (#2) #3}
\def\preprint#1#2 {Preprint #1 #2}
\def\prl#1#2#3  {{\em Phys. Rev. Lett.} {\bf#1} (#2) #3}
\def\prs#1#2#3  {{\em Proc. Roy. Soc.} {\bf#1} (#2) #3}
\def\ptp#1#2#3  {{\em Prog. Th. Phys} {\bf#1} (#2) #3}
\def\ps#1#2#3   {{\em Phys. Scr.} {\bf#1} (#2) #3}
\def\rmp#1#2#3  {{\em Rev. Mod. Phys.} {\bf#1} (#2) #3}
\def\rpp#1#2#3  {{\em Rep. Prog. Phys.} {\bf#1} (#2) #3}
\def\sci#1#2#3  {{\em Science} {\bf#1} (#2) #3}
\def\sjnp#1#2#3 {{\em Sov. J. Nucl. Phys.} {\bf#1} (#2) #3}
\def\spj#1#2#3  {{\em Sov. Phys. JETP} {\bf#1} (#2) #3}
\def\spu#1#2#3  {{\em Sov. Phys.-Usp.} {\bf#1} (#2) #3}
\def\zp#1#2#3   {{\em Zeit. Phys.} {\bf#1} (#2) #3}
\def\gev{\,{\rm GeV}}
\def\mev{\,{\rm MeV}}
\def\mpc{\,{\rm Mpc}}
\def\tev{\,{\rm TeV}}
\begin{document}
\pagestyle{empty}
\begin{flushright}
{\tt CERN-TH.95/134\\ OUTP-95-15P\\ hep-ph/9506283\\ (revised)}
\end{flushright}
\vspace{15mm}
\begin{center}
{\Huge Successful Supersymmetric Inflation}\\
\vspace{15mm}
{\Large G.G. Ross~$^\diamondsuit$\footnote{SERC Senior Fellow, on leave
                                             from $\heartsuit$} \ and \ 
        S. Sarkar~$^\heartsuit$\footnote{PPARC Advanced Fellow}}\\
\vspace{10mm}
$^\diamondsuit${\sl Theory Division, C.E.R.N., CH-1211, Geneva 23, 
                     Switzerland} \\
$^\heartsuit${\sl Theoretical Physics, University of  Oxford,\\ 
                   1 Keble Road, Oxford OX1 3NP, U.K.}\\
\end{center}
\vspace{10mm}

\begin{abstract} 
  We reconsider the problems of cosmological inflation in effective
  supergravity theories. A singlet field in a hidden sector is
  demonstrated to yield an acceptable inflationary potential, without
  fine tuning. In the simplest such model, the requirement of
  generating the microwave background anisotropy measured by COBE
  fixes the inflationary scale to be about $10^{14}\gev$, implying a
  reheat temperature of order $10^{5}\gev$. This is low enough to
  solve the gravitino problem but high enough to allow baryogenesis
  after inflation. Such consistency requires that the generation of
  gravitational waves be negligible and that the spectrum of scalar
  density perturbations depart significantly from scale-invariance,
  thus improving the fit to large-scale structure in an universe
  dominated by cold dark matter. We also consider the problems
  associated with gravitino production through inflaton decay and with
  other weakly coupled fields such as the moduli encountered in
  (compactified) string theories.
\end{abstract}

\vspace{1cm}
\begin{center}
Accepted for publication in {\sl Nuclear Physics B}
\end{center}

\newpage
\pagestyle{plain}
\section{Introduction}

Although inflation~\cite{guth81} is an attractive solution to the
horizon/flatness problems of the standard Big Bang model and to the
cosmological monopole problem of GUTs, it has yet to find a compelling
physical basis~\cite{books}. Interest in this question has been
rekindled by the COBE~\cite{cobe92} discovery of temperature
fluctuations in the cosmic microwave background (CMB) consistent with
a `Harrison-Zeldovich' scale-invariant power spectrum. This arises
naturally in `slow-roll' inflationary models from quantum fluctuations
of the scalar field which drives the de~Sitter phase of exponential
expansion, as it evolves towards its minimum~\cite{fluc}. The observed
small amplitude, ${\delta}T/T\sim10^{-5}$, requires an extremely flat
scalar potential stabilized against radiative corrections. This picks
out a gauge singlet field in a theory incorporating supersymmetry
(SUSY) as the most likely candidate for the
`inflaton'~\cite{susyinfl,hrr84,susyinflrev}.

However such models contain very weakly coupled fields having masses
of $\O(m_W)$ and this creates difficulties with the cosmological
history {\em after} inflation. For example, gravitinos can have
observable effects on the standard cosmology since they decay very
late with lifetime $\sim~M_{\P}^2/m_{3/2}^3$~\cite{wein82}. Although
their primordial abundance can be inflated away, they are recreated
during `reheating' as the inflaton oscillates about its minimum,
converting vacuum energy into radiation~\cite{ekn84}. This imposes a
severe constraint on the reheat temperature since even a small number
of massive late decaying particles can disrupt primordial
nucleosynthesis or the thermalization of the
CMB~\cite{constraints}. Also inflation offers no obvious solution to
the `Polonyi problem'~\cite{polprob,inflmass} which is rather more
subtle, being associated with the production of the unwanted states
via vacuum decay in sectors whose vacuum energy is affected through
gravitational interactions by the vacuum energy driving
inflation. This is of particular relevance to the moduli in
(compactified) string theories~\cite{moduliprob}.

In this paper we introduce and discuss several novel features of
inflation in superstring theories with moduli fields. The central
question is whether acceptable inflation can occur in a natural way
with no need for fine tuning or for multiple correlated periods of
inflation. We argue that this is indeed the expectation in the
presence of moduli fields. First we consider the possible scale for an
inflationary potential in theories with a single stage of dynamical
symmetry breakdown responsible for SUSY breaking in a `hidden
sector'. Contrary to commonly expressed opinions, it appears quite
plausible for the scale of the inflationary potential to be of
$\O(10^{14})\gev$, consistent with the value needed to generate the
observed CMB fluctuations. We then discuss whether fine tuning is
neccessary in order to achieve a sufficiently flat potential. The
novel feature considered here is the possibility that the initial
conditions of moduli fields selected with just the values needed to
generate successful inflation. This happens because the moduli fields
are expected to have chaotic initial values, hence there will be {\em
some} domains with the right initial conditions to give an
inflationary potential. Such domains will inflate leading, through
quantum fluctuations, to `eternal' inflation; these configurations
will thus dominate the final state of the universe. With this
motivation we consider the resultant inflationary potential, using a
simple example as a guide, and find that it has a significant
curvature, at about the maximum possible for slow roll inflation.

We then consider the reheating phase and, in particular, the
post-inflation production of gravitinos or other weakly coupled
states. Demanding that their thermal production during reheating be
under control implies a general constraint on the inflationary
potential and leads to the observationally testable prediction that
there should be a negligible tensor component in the CMB anisotropy in
models with a single stage of inflation. A second prediction following
from the curvature of the inflationary potential is that the spectrum
of scalar density perturbations should deviate from scale-invariance
so as to suppress small-scale power. This provides a better fit to the
observed clustering and motions of galaxies in a cold dark matter
(CDM) universe. We demonstrate how even direct gravitino production is
acceptably small in the favoured class of inflationary models in which
the inflaton resides in a hidden sector of the theory. Finally we
discuss the implications of the Polonyi problem, in particular for
recent models which treat moduli as dynamic variables at the
electroweak scale.

\section{Supersymmetric Inflation}

If there is a stage of inflation in the early universe driven by
approximately constant vacuum energy, then requiring that the
associated quantum fluctuations should not create anisotropies in the
CMB in excess of those observed by COBE bounds the scale of this
potential energy, $V^{1/4}$, to be at least two orders of magnitude
 below the Planck scale.\footnote{The generation of gravitational waves
 becomes increasingly pronounced as the inflationary energy scale
 approaches the Planck scale~\cite{fluc}. Assuming these to be entirely
 responsible for the observed quadrupole anisotropy yields the
 conservative bound
 $V^{1/4}\leqsim4\times10^{-3}M_{\P}$~\cite{gravlim}. If the
quadrupole is in fact due to scalar density fluctuations then the
bound is even more restrictive as we shall see.} At this energy scale,
gravitational interactions and interactions due to string and
Kaluza-Klein states with masses of $\O(M_{\P})$ are small, hence one
can use an effective field theory to describe the inflation sector.
This effective theory should of course contain the Standard
$SU(3)_{\c}\otimes\,SU(2)_{\rm L}\otimes\,U(1)_{Y}$ Model. The condition
that the electroweak symmetry breaking scale should naturally be small
($\ll\,M_{\P}$) without fine tuning the parameters of the theory then
requires it to be supersymmetric with SUSY broken (softly) only just
above the electroweak scale. We shall consider in particular
supergravity (SUGRA) theories based on local supersymmetry which
include a description of gravity and are of the type that descend from
the superstring. The $N=1$ SUGRA theory describing the interaction of
gauge singlet superfields is specified by the K\"{a}hler potential
$K(\Phi,\Phi^{\dagger})$, which determines the form of the kinetic
term in the Lagrangian. The scalar potential is~\cite{sugra}
\begin{equation} 
 V = \frac{1}{4} 
  \e^{K} \left[G_{a} (K^{-1})^{a}_{b} G^{b} - 3 \mid{W}\mid^2\right]\ ,
\label{V}
\end{equation}
where the K\"{a}hler function $G_{a}=K_{a}W+W_{a}$, the indices $a,~b$
denote derivatives with respect to the chiral superfields, $\Phi$, and
$W(\Phi)$ is the superpotential which specifies the Yukawa couplings
of the theory and the scalar couplings related by supersymmetry.

\subsection{``Natural'' Inflation}

A criticism that is often levelled at models of vacuum energy driven
inflation is that they require unnatural
fine-tuning~\cite{books}. There are two parts to this problem. First
of all, such inflationary models necessarily contain small parameters
in order to generate an inflationary potential significantly lower
than the Planck scale. To provide a physical justification for this,
one should perhaps associate the inflationary scale with one of the
scales needed in viable unified field theories of the fundamental
interactions. The second aspect of the problem concerns the initial
conditions, specifically the question of how probable are the initial
field configurations necessary to ensure an inflationary era. In this
section we discuss both these issues in the context of supersymmetric
inflation.

\subsubsection{The inflationary scale\label{inflscale}}

Let us first consider the mass scale(s) to be expected in the
effective field theory which may play a role in setting the energy
scale for the potential in the inflation sector. We know that in
string and/or compactified theories there are numerous states with
mass of $\O(M)$, where we use as our basic mass unit $M \equiv
M_{\P}/\sqrt{8\pi}\simeq2.44\times10^{18}\gev$. In addition, the
success of the supersymmetric unification prediction relating the
strong and electroweak gauge couplings suggests there is a stage of
spontaneous gauge symmetry breaking at a scale
$M_{X}\approx2\times10^{16}\gev$~\cite{susyunif}. The theory must also
have a source of SUSY breaking characterized by the gravitino mass,
$m_{3/2}\leqsim1\tev$. The most plausible origin for the latter is
dynamical supersymmetry breaking via nonperturbative effects driven by
a new interaction which becomes strong at a scale $\Lambda_{\c}$, thus
inducing gaugino condensation in a hidden sector. In this case
$m_{3/2}\sim\langle\lambda\lambda\rangle/M^2$ and the gaugino
condensate $\langle\lambda\lambda\rangle\approx(10^{13}\gev)^3$. How
could such scales affect a gauge singlet sector? Consider first the
gauge symmetry breaking scale. We denote by $\chi,\bar{\chi}$, the
gauge non-singlet fields which acquire a vacuum expectation value
(vev) of $\O(M_{X})$ along a $D$-flat direction thus breaking the
gauge symmetry, and by $\Theta,\;\bar\Theta$, the gauge singlet fields
with masses of $\O(M)$. Allowing for a coupling between these fields,
consider a superpotential of the form (suppressing coupling constants
of order unity)
\begin{equation} 
 W = M \Theta \bar\Theta - \bar\Theta \chi \bar{\chi}\ , 
\label{W} 
\end{equation} 
from which we see that the gauge symmetry breaking vev in
$\chi,\bar{\chi}$ induces a vev in the massive field given by
$\langle\Theta\rangle \equiv\Delta
=\langle\chi\rangle\langle\bar{\chi}\rangle/M\approx10^{14}\gev$. If
$\Theta$ couples to other fields in the theory, they will acquire mass
determined by this vev, the value of the mass determined by the
strength of coupling. 

The point is that once a stage of symmetry breaking is generated below
the Planck scale in some sector of the theory, one may expect it to
generate masses in every sector of the theory. However this conclusion
may not be true during an inflationary era in a string theory. The
reason is that in theories of this type all scales other than the
Planck scale must arise dynamically. The most obvious way is through
the supersymmetry breaking sector via non-perturbative effects, which
become strong at a scale $\Lambda_{\c}$ far below the Planck scale,
driving a gaugino condensate. However one must then require that all
other related scales are consistent with the formation of the
condensate. For example if a gauge symmetry breaking vev is generated
it is due to a supersymmetry breaking trigger. A vev for the $\chi$
field introduced above may be induced through radiative breaking in
which the $\chi$ soft-SUSY-breaking mass-squared term in the
Lagrangian is driven negative by radiative corrections at some scale
$M_X$ and minimization of the effective potential for the
$\chi,\bar{\chi}$ fields and the SUSY breaking sector together,
requires the vev to be of $\O(M_X)$, essentially independent of the SUSY
breaking trigger. However this only happens when it is energetically
favourable for the vev to develop and in particular {\em cannot} occur
if there is a large potential energy associated with the inflation
sector. In other words, the inflationary potential will cause $\chi$
to acquire a soft mass term inhibiting the development of its vev
until inflation is over. These considerations place an {\it upper}
limit on the scale of inflation, viz. the positive inflationary
potential should not exceed the energy scale involved in gauge and
SUSY breaking otherwise the latter will be inhibited by the very
inflationary phase it is supposed to drive.

In the literature (e.g. ref.\cite{modulicosm}) this constraint has
been implicitly interpreted as requiring that the inflationary vacuum
energy should not exceed the non-zero $F$- or $D$-term contribution to
the effective potential which is responsible for SUSY breaking. This
is given by $(m_{3/2}M_{\P})^{2}\sim\Lambda_{\c}^6/M^2$, leading to an
upper bound of $\O(10^{11})\gev$ for the scale of the inflationary
potential, which is indeed too small.  There are two possible ways out
of this apparent impasse. It may be that the interaction in a second
sector also becomes strong and dynamically generates a second, larger
scale which is responsible for inflation but does not lead to SUSY
breaking after inflation.  We disfavour this possibility because we
are trying to avoid multiple correlated sectors when constructing a
``natural'' theory. However it is quite reasonable to expect that a
single scale of dynamical SUSY breaking has a {\em different}
magnitude during inflation than after inflation. For example the
inflationary potential may inhibit gauge symmetry breaking in the
manner discussed above. If this symmetry breaking relates to the group
which drives SUSY breaking (when its interaction becomes strong), the
result is that the $\beta$-function of the group is bigger during
inflation, hence the scale at which the coupling becomes strong is
higher, in turn driving a larger scale of SUSY breaking.\footnote{Here
we are assuming that the initial value of the gauge coupling set by
the dilaton vev is fixed, say at a modular invariant point. There is,
of course, the possibility that the dilaton vev also varies, its
potential affected by the inflationary potential. This is another
possible source for a change in the SUSY breaking scale before and
after inflation.}

A second possibility is that the inflation scale is related to the
scale, $\Lambda_{\c}$, at which the coupling becomes large and not to
the final SUSY breaking scale. Indeed this seems to us the more
reasonable choice for certainly $\Lambda_{\c}$ is a relevant scale of
new physics, for example, connected with the appearance of confined
massive states. Thus it is perfectly possible that due to the initial
conditions the vacuum energy starts off at the natural binding energy
scale $\Lambda_{\c}^4$ and only drops to the scale
$\Lambda_{\c}^6/M^2$ as the fields (in this case the inflaton) adjust
to their vacuum values. Clearly such a scenario requires a strong
cancellation between the terms contributing to the vacuum energy at
the minimum but this is just what is usually assumed about gaugino
condensation. As we shall demonstrate, inflation at the
$\O(10^{14})\gev$ scale is compatible with all constraints for a
sub-class of inflationary models which are of the
`new'~\cite{new,linde2} rather than of the `chaotic'~\cite{chaotic}
type.\footnote{In order to realise the latter, one must rely on a
small coupling constant (rather than a ratio of mass scales) to
provide the scale of inflation in terms of the Planck scale. In string
theories small couplings can indeed arise but only dynamically when
moduli fields acquire large vevs. These however do not lead to
inflationary potentials because the would be inflationary potential is
not sufficiently flat in the moduli direction~\cite{stringinfl}.} This
is marginally consistent with the identification of the scale with
$\Lambda_{\c}\sim10^{13}\gev$, given the uncertanties involved in this
identification.

Within these constraints we consider it to be the expectation rather
than the exception for there to be sectors in the theory associated
with the mass scale required for acceptable inflation. To demonstrate
this we may use the mechanism discussed above to construct a model for
inflation in which the scale of the potential is set by the original
spontaneous breaking of the gauge symmetry (and, if this scale is not
fundamental, by the associated SUSY breaking trigger, provided the
scale is constrained in the manner just discussed). The starting point
is the form of the potential describing the inflaton which, for the
reasons discussed above, we take to be a gauge singlet, $\phi$; at
this stage we need not specify whether it is an elementary scalar
field or arises as a composite object. The coupling between the
superfield $\Phi$ containing the inflaton and $\Theta$ will be
constrained by any symmetries of the theory. The superpotential of
eq.(\ref{W}) has an $R$-symmetry under which $\Theta$ and
$\chi\bar{\chi}$ transform as $\e^{i\gamma}$, $\bar\Theta$ transforms
as $\e^{i(2\beta-\gamma)}$ while the superspace coordinate transforms
as $\e^{-i\beta}$. Consider the sector with an $R$-singlet, gauge
singlet, superfield $\Phi$ which contains the (complex) inflaton
field, $\phi$ as its scalar component. The most general
superpotential, $P$, describing $\Theta,\;\bar\Theta$, $\chi$ and
$\Phi$, has the form dictated by this $R$-symmetry,
\begin{equation} 
 P = \Theta \bar\Theta M\ f\left(\frac{\Phi}{M}\right) 
     + \bar\Theta \chi\bar{\chi}\ , 
\label{P} 
\end{equation} 
where $f(x)$ is a function which is not constrained by the
$R$-symmetry alone. (We have absorbed the constant term generating the
$\Theta$ mass in $f$.) As discussed above, once $\chi,\bar\chi$
acquire vevs breaking the gauge symmetry, the field $\Theta$ will also
acquire a vev,
\begin{equation}
 \Delta = \frac{\langle\chi\rangle\langle\bar{\chi}\rangle}{2M}\ ,
\label{Delta}
\end{equation}
(with $f(0)=1$) leading to the inflaton superpotential
\begin{equation}
 I\,(\Phi) = \Delta^2 M\ f\left(\frac{\Phi}{M}\right). 
\label{I} 
\end{equation} 
Of course there are other ways to motivate this form; the most obvious
is that it arises directly from the gaugino condensate sector in which
the confined states have mass of $\O(\Delta)$ and the inflaton is one
of these composite states.  We shall consider a particular choice for
$f(\Phi/M)$ in Section~\ref{simple} but first we consider the general
features of such a superpotential. The first point is that the
associated potential for the inflaton $\phi$ is proportional to
$\Delta^4$, showing how the gauge breaking scale naturally sets the
scale for the potential of light fields in the effective potential
obtained by integrating out massive fields. By appealing to a richer
symmetry structure one can construct examples in which the scalar
potential is proportional to different powers of $\Delta$, but this
example suffices to illustrate the general point. Moreover it will
turn out that this power is just what is required for acceptable
inflation generating the observed amplitude of density perturbations.

\subsubsection{Initial conditions}

We turn now to the second part of the naturalness question, namely
whether one requires fine tuning of the initial conditions or of the
parameters determining the potential for there to be an inflationary
era~\cite{initcond}. Consider the structure of the scalar potential
following from the superpotential, $I$. It will be determined once the
K\"{a}hler potential is specified but for now we need merely assume
that it has at least one local maximum or point of inflection. If the
initial conditions are such that the field lies very close to this
point then there will be an inflationary era.  Of course if the
initial value of the field is determined by the high temperature
effective potential~\cite{susytemp} then it is unlikely that this
value coincides with the point at which the first derivative of the
potential vanishes, unless one fine tunes the parameters of the
potential to ensure this. If, however, thermal effects do not fix the
initial state, then it will have, in general, a broad
distribution~\cite{chaotic,initcond}. In this case, while there may
only be a small probability that one starts with the value needed for
an inflationary era, the region posessing this initial value {\em
will} inflate and become the overwhelmingly probable state after
inflation. This is the basic idea which leads us to the conclusion
that weakly coupled fields which are not in thermal equilibrium below
the Planck scale are overwhelmingly likely to generate inflation
without the need for any fine tuning.

To quantify this, consider a field $\phi$ which drops out of thermal
equilibrium at temperatures below the Planck scale. (Note that this is
what we expect for the field with interactions described by the
superpotential of eq.(\ref{I}) because the leading cubic term giving a
Yukawa coupling and associated quartic scalar couplings, has a
coupling of $\O(\Delta^2/M^2)$.) At the Planck temperature it is
reasonable to suppose that through gravitational effects the field has
thermal fluctuations given by $\langle(\delta\phi)^2\rangle\sim~M^2$
about some value set by the physics at the Planck scale.\footnote{It
 has been argued~\cite{chaotic} that a reasonable initial condition
 emerging from the physics at the Planck scale is a flat probability
 distribution for values of the field which have $V(\phi){\leqsim}M^4$.
 For a generic potential $V=\lambda\phi^4$ with small coupling
 $\lambda$, the field value can be thus much larger than the
 Planck scale as is required in chaotic inflation. However in string
 theories couplings are of order unity so the initial field value
 should {\em not} be taken to be much larger than the Planck scale.}
Due to the spatial derivative terms in the Lagrangian the field will
be smoothed out over a horizon volume by the Kibble mechanism. Thus as
the universe cools below the Planck scale, one might suppose that the
fields free stream and (averaging over the enlarged horizon) settle in
to their Planck scale mean values with deviation $\propto~M/\sqrt{n}$
at a temperature where the horizon contains $n$ Planck volumes.
However as the temperature drops the potential energy associated with
the random initial conditions will become comparable to the thermal
energy and the field expectation values will move to reduce the
potential. This will be a complicated process, dependent on the
initial field configurations, in which the various fields do {\em not}
free stream but are correlated in any given horizon volume and as a
result the fields are {\em not} likely to head to the mean values
fixed at the Planck scale. Thus we conclude that the field values in
different horizon volumes will, apart from sharing excluded regions of
field space for which the potential energy is greater than the thermal
energy, be largely uncorrelated until a (very low) temperature at
which the excluded regions of field space grow sufficiently precisely
to specify the allowed regions. The implication is that one should use
a (flat) distribution of initial values, as at the Planck scale.

Let us now consider the case when the mean value of the field in any
horizon volume lies at a local maximum or point of inflection of the
potential, at the temperature for which the potential energy at this
point is comparable to the thermal energy. This is just the situation
discussed in ref.~\cite{gp85} wherein it was shown that despite the
fluctuations of the field throughout the horizon volume it behaves as
if it were a classical field with value given by the mean and evolving
according to the equation of motion
\begin{equation}
 \ddot \phi +3 H \dot \phi = -\frac{\partial V}{\partial \phi} 
       \approx m_{\phi}^2\phi\ ,
\end{equation} 
where we have expanded about the point of the potential for which the
first derivative vanishes. Inflation will occur if the first term on
the LHS is negligible (i.e. if $m_{\phi}<H$). Since the two quantities
are typically of the same order, whether this happens or not depends
on the detailed values of the parameters. Assuming that it does,
inflation will continue until the curvature of the potential becomes
significant at $\phi\sim~M$. This determines the number of e-folds of
inflation to be
\begin{equation}
 N = H t_{\rm inflation} \sim \frac{3\Delta^4}{M^2 m_{\phi}^2}
     \ln\left(\frac{M}{\phi_0}\right),
\label{N} 
\end{equation}
where $\phi_0$ is the initial value of $\phi$. This is given by
$\phi_0\sim~T_{\deS}$ the temperature at which the thermal and false
vacuum energy in the de~Sitter phase are about equal~\cite{gp85}; thus
$T_{\deS}\sim\Delta$, which can be used in eq.(\ref{N}) to determine
$N$. If there are no very small parameters we have $m_{\phi}\sim~H$,
so inflation will be limited to a few tens of e-folds. Now we can
complete the quantitative estimate of the probability that we end up
with an inflationary universe. With our best guess for the initial
conditions being an uniform distribution over the Planck scale, the
probability we start with $\phi$ within a distance $\phi_0$ of the
maximum of the potential is just $\phi_0/M\sim\Delta/M\sim10^{-4}$.
However this is more than compensated for by an inflation of tens of
e-folds, so the final state of the universe is dominated by states
which have enjoyed an inflationary era.

As we have seen, the initial field configuration with vanishing
derivative for the scalar potential dominates the final state of the
universe due to the inflationary expansion associated with such a
configuration. One may then consider the possibility that the second
derivative also vanishes, for in this case the amount of inflation
increases. In theories following from the superstring such a
possibility appears to be quite likely because the couplings which
determine the higher derivatives are themselves determined by the vevs
of moduli fields. If the initial values of these vevs are random,
there will be regions in which the initial conditions are such that
the second derivative vanishes too. This process will not continue to
an arbitrary degree for there are only a finite number of moduli and
the condition that a higher derivative vanishes determines some
combination of them. However it is quite plausible that there are
sufficient moduli~\footnote{In Calabi-Yau compactification there is
one complex structure moduli for each (2,1) form and one K\"{a}hler
moduli for each (1,1) form.} undetermined by the time the inflationary
potential dominates that at least the second derivative
vanishes. (Then there will be many more e-folds of expansion than the
$\sim50-60$ needed for the observable universe to arise from one
causally connected domain.) While it is possible that further
derivatives may vanish, this is irrelevant in determining the
implications for the observable universe, so in what follows we
consider examples in which only the first two derivatives are zero.

In fact it has been pointed out to us (A.~Linde, private
communication) that this picture fits in very well with the concept of
`everlasting eternal inflation'~\cite{llm94} in which, during the
slow-roll, quantum fluctuations continally produce regions {\em
closer} to the maximum of the potential. Although the initial
probability for this is small, these regions become exponentially
larger than the remainder and thus ultimately dominate the final state
of the universe. In this way the inflationary universe reproduces
itself eternally, so essentially all final states descend from an
inflationary era. In terms of the moduli, the initial state is
continually being driven to the optimal choice of moduli vevs for
which the infationary potential is flattest. This is the ultimate
``fine tuning without fine tuning''! We note that for {\em any}
inflationary model it is essential to invoke a mechanism of this sort
if one is to overcome the argument\cite{penrose} that there is
apparently a hopelessly small chance that the initial conditions of
the universe are sufficiently homogeneous and isotropic for inflation
to start at all.

\subsection{Constraints on the inflationary potential}

Before making a specific choice for $f(\Phi/M)$ in eq.(\ref{I}) let us
consider the overall constraints on the scale of inflation. In order
to illustrate these, we specialize to the minimal choice of K\"{a}hler
potential, $K=\Phi^\dagger\Phi$, corresponding to canonical kinetic
energy for the scalar fields. (This is known to be the case in certain
orbifold models, however here we make this choice for pedagogic
reasons; the discussion can be extended to non-minimal K\"{a}hler
potentials as well.) With this choice the scalar potential following
from the superpotential, $I$, is~\cite{sugra}
\begin{equation} 
 V_{I} (\phi) = \e^{|\phi|^2/M^2}
  \left[\left|\frac{\partial I}{\partial \phi} 
  + \frac{\phi^*I}{M^2}\right|^2 - \frac{3 |I|^2}{M^2}\right]_{\Phi=\phi}.
\label{VI} 
\end{equation}
In what follows we shall consider slow-roll models in which the
function $f$ can be expanded as a power series in $\phi/M$ or, as in
simple chaotic inflation models where $\phi/M$ is large, is given by a
combination of powers. If there are no small parameters other than
$\Delta/M$ (as argued above, this is the simplest case and may
eliminate the need for fine tuning) then $V$ has a minimum at
$\phi\sim~M$. Clearly more complicated models can be constructed and
we will comment on these later but our purpose here is to discuss
whether the bounds on the reheat temperature imposed by consideration
of the production and subsequent decay of unstable gravitinos present
a problem for the generic model characterized by a {\em single} energy
scale.

\subsubsection{The gravitino problem}

Gravitinos or other unwanted fields will be thermally excited during
the reheating epoch following inflation and decay subsequently into
high energy particles. The effects of such decays on cosmological
observables such as the primordial abundances of helium and deuterium
and the spectrum of the CMB have been studied in some detail. The most
stringent constraint on the gravitino abundance comes from the
observational upper limit on the abundances of D and $^{3}$He which
can be produced via photofission of $^{4}$He by the radiation cascades
from gravitino decay:
$m_{3/2}(n_{3/2}/n_{\gamma})\leqsim3\times10^{-12}\gev$~\cite{ens85}.
Taking the gravitino abundance produced during reheating by $2\to2$
processes involving gauge bosons and gauginos to be~\cite{ekn84}
\begin{equation}
 \frac{n_{3/2}}{n_{\gamma}} \simeq 2.4\times10^{-13} 
  \left(\frac{T_{\R}}{10^9\gev}\right),
\label{ekn}
\end{equation}
this implies an upper bound on the reheat temperature of~\cite{ens85}
\begin{equation}
 T_{\R} \leqsim 2.5\times10^8 \gev \quad 
  \for \quad m_{3/2} = 100\ \gev.
\end{equation} 
Subsequent work~\cite{constraints} has shown that the true constraint
is actually more restrictive than in ref.\cite{ens85} by a factor upto
$\sim2$ for a radiative lifetime greater than $\sim2\times10^7\sec$
(corresponding to $m_{3/2}\leqsim300\gev$) but less stringent for
shorter lifetimes. Here we have taken the lifetime for gravitino decay
into a gauge boson-gaugino pair,
$\tau_{3/2}\sim4M_{\P}^2/N_{\c}m_{3/2}^3$ where $N_{\c}$ is the number
of available channels (assuming
$m_{\tilde{\gamma},\tilde{g}}\ll{m_{3/2}}$), so that~\cite{ekn84}
\begin{eqnarray}
 \tau_{3/2 \to \tilde{\gamma} \gamma} &\simeq 3.9\times10^{5} \sec
  &\left(\frac{m_{3/2}}{\tev}\right)^{-3}, \nonumber\\
 \tau_{3/2 \to \tilde{g} g} &\simeq 4.4\times10^{4} \sec
  &\left(\frac {m_{3/2}}{\tev}\right)^{-3}.
\end{eqnarray}
For a gravitino mass of $1\tev$, the radiative lifetime is about
$4\times10^5\sec$ and the relevant constraint now comes from requiring
that the photofission of deuterium not reduce its abundance below the
observational lower limit~\cite{jss85,dehs89}. A careful calculation
shows that this requires
$m_{3/2}(n_{3/2}/n_{\gamma})\leqsim5\times10^{-10}\gev$~\cite{constraints}
so the reheat bound is weakened to $T_{\R}\leqsim2\times10^{9}\gev$.
Photofission processes become ineffective for $\tau\leqsim10^4\sec$
but now there are new constraints from the effect of hadrons in the
showers on the $^4$He abundance. If the gravitino mass is 10~TeV with
a corresponding lifetime of $\tau_{3/2\to\tilde{g}g}\sim50\sec$, this
constraint is
$m_{3/2}(n_{3/2}/n_{\gamma})\leqsim1.5\times10^{-8}\gev$~\cite{hadro},
hence $T_{\R}\leqsim6\times10^9\gev$. There is {\em no} bound on
$T_{\R}$ for gravitinos of mass exceeding $50\tev$ which decay before
nucleosynthesis~\cite{hadro,wein82} but such a large mass cannot be
accomodated in minimal models. Hence we take the reheat bound
corresponding to the relic abundance in eq.(\ref{ekn}) as given in
ref.\cite{constraints},
\begin{eqnarray}
 T_{\R} &\leqsim &10^{8},\ 2\times10^9,\ 6\times10^{9}\ \gev\ ,
  \nonumber\\ 
  \mbox{for} \quad m_{3/2} &= &10^2, \qquad 10^3,\qquad 10^4\ \gev\ .
\label{conTR}
\end{eqnarray}
We emphasize that these are {\em conservative} bounds. For example in
ref.\cite{km94} the relic gravitino abundance is calculated to be a
factor of $\sim4$ higher than in eq.(\ref{ekn}) after including
interaction terms between the gravitino and chiral multiplets; this
would tighten all the above bounds on $T_{\R}$ by the same factor.
However, the authors of ref.\cite{km94} also numerically calculate the
cosmological constraint based on D+$^{3}$He photoproduction to be,
inexplicably, a factor of $\sim30$ more stringent than the one
calculated analytically in ref.\cite{constraints}, which was confirmed
by the full Monte Carlo simulation of ref.\cite{psb95}. Consequently
they obtain bounds on the reheat temperature which are much more
severe than in eq.(\ref{conTR}), for example
$T_{\R}\leqsim2\times10^{6}\gev$ for $m_{3/2}\sim100\gev$.

Recently the thermal production of gravitinos has been
reexamined~\cite{fischler}; it is argued, contrary to previous
studies~\cite{ekn84}, that massive gravitinos {\em can} achieve
equilibrium with the background plasma at temperatures well below the
Planck scale via interactions of their longitudinal spin-1/2
(goldstino) component with a cross-section which increases as $T^2$
due to the breaking of SUSY by finite temperature effects.  The relic
abundance is then given by~\cite{fischler}
\begin{equation}
 \frac{n_{3/2}}{n_{\gamma}} \approx
  \frac{g_*^{1/2}\alpha_{\s}^3 T^3}{m_{3/2}^3 M_{\P}^2} \quad
  \sim 3\times10^{-13} \left(\frac{T_{\R}}{10^5\gev}\right)^3
            \left(\frac{m_{3/2}}{\tev}\right)^{-2} ,
\end{equation}
where $g_*$, the number of relativistic degrees of freedom, equals
915/4 in the minimal supersymmetric standard model (MSSM) at
$T\gg{m_Z}$. Imposing the nucleosynthesis constraints discussed above
then yields a very restrictive upper bound on the reheat temperature:
\begin{eqnarray}
 T_{\R} &\leqsim &2\times10^{4},\ 10^6,\ 2\times10^{7}\ \gev\ , 
  \nonumber\\ 
  \mbox{for} \quad m_{3/2} &= &10^2, \qquad 10^3,\qquad 10^4\ \gev\ .
\label{newTR}
\end{eqnarray}
However this claim has been questioned on general grounds~\cite{lr95}
and an explicit calculation~\cite{enor95} of finite-temperature
effects demonstrates that these do not alter the estimate in
eq.(\ref{ekn}). Hence we will consider the impact on inflationary
models of the reliable bound~(\ref{conTR}) but also note the
implication of the proposed new bound~(\ref{newTR}), assuming a
nominal gravitino mass of 1 TeV.

Another aspect of the `gravitino problem' is the need to avoid direct
production of gravitinos from the decay of the inflaton field during
the reheating process~\cite{hrr84}, given that there is always a
gravitational strength coupling between the gravitino (or any other
state) and the inflaton. Unlike thermal production discussed above,
this needs to be considered in the context of a specific model and we
shall do so in Section~\ref{direct}.

\subsubsection{Normalization to COBE}

The semi-classical equation of motion for the inflaton field
is~\cite{books}
\begin{equation}
 \ddot{\phi} + 3 H \dot{\phi} + V' (\phi) = 0.
\label{evolve}
\end{equation}
Essentially all models generating an exponential increase of the
scale-factor $a$ satisfy the slow-roll conditions~\cite{ll94}
\begin{equation}
 \dot{\phi} \simeq -\frac{V'}{3H}\ , \quad
 \epsilon \equiv \frac{M^2}{2} \left(\frac{V'}{V}\right)^2 \ll 1, \quad
 |\eta| \equiv \left|M^2 \frac{V''}{V}\right| \ll 1.
\label{slowroll}
\end{equation}
Inflation ends (i.e. $\ddot{a}$ becomes zero) when
\begin{equation}
 {\rm max} (\epsilon, |\eta|) \simeq 1. 
\label{end}
\end{equation}
Given these conditions, the spectrum of 
adiabatic density perturbations is~\cite{lyth}
\begin{equation}
 \delta^2_{\H}(k) = \frac{1}{150 \pi^2} \frac{V_{\star}}{M^{4}} 
  \frac{1}{\epsilon_{\star}},
\label{deltah}
\end{equation}
where $\star$ denotes the epoch at which a scale of wavenumber $k$
crosses the `horizon' $H^{-1}$ during inflation, i.e. when $aH=k$.
(Note that for a scale-invariant spectrum $\delta_{\H}$ equals
$\sqrt{4\pi}$ times the parameter $\epsilon_{\H}$,
cf. ref.\cite{ll94}.) The CMB anisotropy measured by COBE allows a
determination of the perturbation amplitude at the scale $k_1^{-1}$
corresponding roughly to the size of the presently observable
universe, $H_0^{-1}\simeq~3000~h^{-1}\mpc$, where
$h\equiv~H_0/100$~km$~\sec^{-1}\mpc^{-1}$ is the present Hubble
parameter. The number of e-folds from the end of inflation when this
scale crosses the horizon is~\cite{ll94}
\begin{eqnarray}
 N_1 \equiv N_{\star} (k_1) &\simeq 51 
 &+ \ln\left(\frac{k_1^{-1}}{3000h^{-1}\mpc}\right)
 + \ln\left(\frac{V_{\star}}{3\times10^{14}\gev}\right)
 + \ln\left(\frac{V_{\star}}{V_{\en}}\right) \nonumber\\
 &&- \frac{1}{3}\ln\left(\frac{V_{\en}}{3\times10^{14}\gev}\right) 
 + \frac{1}{3}\ln\left(\frac{T_{\R}}{10^5\gev}\right),
\end{eqnarray}
where we have indicated the numerical values we anticipate for the
various scales. The exact procedure for normalization is somewhat
subtle~\cite{bsw95} since it depends on the precise shape of the
density perturbation spectrum and on whether there are other
contributions to the CMB anisotropy such as from gravitational waves.
The higher multipoles in the CMB anisotropy also depend on the assumed
composition of the dark matter (i.e. `cold' or `hot') since this
determines how the primordial spectrum is modified on scales smaller
than the horizon at the epoch of matter-radiation equality i.e. for
$k^{-1}<k_{\eq}^{-1}\simeq80~h^{-1}$ Mpc, assuming the dark matter to
contribute the critical density~\cite{cosmofluc}. However, these are
small effects and well within the observational uncertainties so
initially we assume that the anisotropy in the COBE data arises purely
due to the Sachs-Wolfe effect on super-horizon scales
($k^{-1}>k_{\dec}^{-1}\simeq~180~h^{-1}\mpc$) at CMB decoupling. The
best fitting quadrupole moment obtained from the angular power
spectrum of the 2-year data, $Q_{\rms}\simeq~20~\mu$K~\cite{cobe94},
then corresponds to
\begin{equation}
 \delta_{\H} = \sqrt{\frac{48}{5}} \frac{\langle{Q}\rangle}{T_0} 
  \simeq 2.3\times10^{-5}
\end{equation}
with an uncertainty of $\sim10\%$, taking $T_0\simeq2.73$~K. (This is
about $30\%$ higher than the value obtained previously~\cite{ll94}
from the 1-year COBE data~\cite{cobe92}.) Using eq.(\ref{deltah})
then gives
\begin{equation}
 V_{1}^{1/4} \simeq 7.3\ \epsilon_{1}^{1/4}\times10^{16} \gev,
\end{equation}
demonstrating how the inflationary scale is strictly bounded from
above~\cite{ll94}, thus justifying our neglect of gravitational
corrections to the potential. Applied to the supergravity potential
following from eqs.(\ref{I}) and (\ref{VI}) we have
\begin{equation}
 V_{1} = \Delta^{4}\ \e^{|\phi_{1}|^2/M^2} \left[\left| f' 
  \left(\frac{\phi_{1}}{M}\right) + \frac{\phi_{1}^{*}f}{M}\right|^2 
  - 3 |f|^2\right].
\end{equation}
For convenience we have chosen to define $f$ in eq.(\ref{I}) such
that the RHS is just $\Delta^4$, hence the COBE normalization requires
\begin{equation}
 \frac{\Delta}{M} \simeq 3\times10^{-2} \epsilon_{1}^{1/4}.
\label{scale}
\end{equation}

At the end of inflation, the field $\phi$ begins to oscillate about
its minimum until it decays, thus reheating the universe.  In the
class of models considered here, $\phi$ is a gauge singlet field in a
hidden sector with only gravitational strength couplings to other
states. Such models have been widely explored for they can readily
produce an acceptable inflationary potential; clearly this class of
models also has the best chance of satisfying the bound on the reheat
temperature since here the inflaton has the smallest possible coupling
to other states. The dominant coupling of $\phi$ to states $\chi$ in
another sector with superpotential $P(\chi)$ has the form
$\left(\partial{V}/\partial{\phi}\right)~P(\chi)_{A}M^{-2}$, where the
subscript $A$ denotes that the chiral superfields in $P$ should be
replaced by their scalar components. We see that this generates a
trilinear coupling to the light matter fields $\chi$ of strength
$\sim\Delta^2/M^2$, corresponding to a decay width
$\Gamma_{\phi}\sim[m_{\phi}/(2\pi)^3](\Delta^2/M^2)^2$. With our
simplifying assumption that there are no small parameters in $f(\phi)$
the mass of the inflaton is
\begin{equation}
 m_{\phi} \sim \frac{\Delta^2}{M}\ .
\end{equation}
The inflaton thus decays at
\begin{equation}
 t_{\R} \sim \Gamma_{\phi}^{-1} \sim (2 \pi)^3 \frac{M^5}{\Delta^6}
\label{tR}
\end{equation}
and converts its energy content into radiation according to
\begin{equation} 
\rho_{\phi} (t_{\R}) \simeq \frac{\pi^2}{30} g_{*} (T_{\R}) T_{\R}^4\ . 
\end{equation} 
(The `parametric resonance' effect discussed in ref.\cite{reheat} is
irrelevant here since the inflaton has no coupling of the form
$\phi^2\chi^2$ but only terms involving $\chi^3$. This is because the
supergravity couplings involve $P^2$ where $P$ is the superpotential
which is {\em trilinear} in the scalar fields, hence bilinear terms
are suppressed by $m_{\chi}/M_{\P}$.) The temperature at the begining
of the standard radiation-dominated era is thus
\begin{equation} 
T_{\R} \sim \left(\frac{30}{\pi^2 g_{*}}\right)^{1/4} 
 (\Gamma_{\phi} M)^{1/2} \simeq 2.2\times10^{-2}\frac{\Delta^3}{M^2}\ .
\end{equation} 
Demanding that this be less than the phenomenological bounds
(eqs.\ref{conTR},\ref{newTR}) then requires,
\begin{eqnarray}
 \left\{\frac{\Delta}{M}, \epsilon_1\right\} 
  &\leqsim \{2.7\times10^{-4},\ 6.1\times10^{-9}\}
   \quad &\for \quad T_{\R} \leqsim 10^6 \gev, \nonumber\\
  &\leqsim \{3.3\times10^{-3},\ 1.5\times10^{-4}\}  
   \quad &\for \quad T_{\R} \leqsim 2 \times 10^9 \gev,
\label{constrain}
\end{eqnarray}
where we have used eq.(\ref{scale}). Hence the constraint following
from the need to inhibit thermal production of gravitinos forces the
inflationary scale to be low; the COBE normalization then requires the
scalar potential to be very flat indeed. In turn this implies that the
ratio of the tensor to the scalar contributions to the observed CMB
anisotropy, given by $R\simeq~12.4\epsilon_{1}$~\cite{ll94}, is
negligible small. For example, taking the highest allowed value
$\epsilon_1\leqsim1.5\times10^{-4}$ corresponding to the conservative
reheat bound $T_{\R}\leqsim\,2\times10^9\gev$, we have the constraint
\begin{equation}
 R \leqsim 1.9\times10^{-3}.
\end{equation}
This prediction can probably be observationally tested~\cite{grav} by
ongoing and forthcoming medium- and small-scale angular anisotropy
experiments~\cite{wss94}.

We have established the general conditions~(\ref{constrain}) which
ensure a sufficiently low reheat temperature. Can these be achieved in
realistic models? As discussed in Section~\ref{inflscale}, it is
reasonable that a stage of symmetry breaking below the Planck scale
should generate an inflationary potential with a related scale. Our
example of gauge symmetry breaking gave $\Delta\sim3\times10^{-5}M$
(eq.\ref{Delta}), so one can easily obtain the required scale of
inflation without any fine tuning. To discuss the small value required
for the slope of the potential, we will have to specify what the
dynamics of our inflationary scheme is, viz. new or chaotic. In new
inflation~\cite{new}, the inflaton rolls from an initial value near
the origin towards its global minimum at the Planck scale; the
derivative of the potential is neccessarily small in order to ensure
sufficient inflation, hence such models {\em naturally} ensure a small
value for $\epsilon_{1}$. However in chaotic inflation~\cite{chaotic},
the inflaton starts rolling towards its minimum at the origin from an
initially large value beyond the Planck scale.  Hence for a generic
power-law potential $V\propto\phi^\alpha$, $MV'/V$ can be small only
if the value of $\phi$ is much larger than the Planck scale. The field
value corresponding to the spatial scales probed by COBE is
$\phi_{1}\simeq(2\alpha~N_1)^{1/2}~M$, so the slope at this point is
\begin{equation} 
 \epsilon_{1} \simeq \frac{\alpha}{4N_1}\ . 
\end{equation} 
This is $\sim2\times10^{-2}$ for $V\propto\phi^4$, hence violates even
the conservative bound in eq.(\ref{constrain}) by a factor of
$\sim100$. We conclude that it is not possible to satisfy the
phenomenological constraints on the reheat temperature in any
(one-scale) chaotic supergravity inflation model. (Conversely, such
models predict a substantial tensor component in the CMB anisotropy,
e.g. $R\simeq25\%$ for a quartic potential.) We emphasize that by
``chaotic'' we specifically mean here models where the scalar field
vev {\em decreases} during inflation from an initially large value
beyond the Planck scale. We are not referring to the random initial
conditions of chaotic models, which we do indeed adopt ourselves.

So far our analysis has been rather general. We now consider a
specific example capable of giving inflation with small $\phi/M$,
which naturally satisfies the bound on the reheat temperature when the
scalar density perturbations are normalized to the COBE observations.
We will be able to study the remaining questions related to the reheat
process in the context of this hybrid model which combines the
dynamical evolution of new inflation with the initial conditions of
chaotic inflation.\footnote{Such a hybrid model was first studied in
ref.\cite{linde2}.}

\section{A simple model\label{simple}}

In this model~\cite{hrr84} the inflation sector $I$, the
SUSY breaking sector $S$ and the visible sector $G$ interact
with each other only gravitationally and hence may be constructed
separately in the superpotential. First consider the inflation
sector. Requiring that SUSY remain unbroken in the global
minimum, i.e.
\begin{equation}
 \left|\frac{\partial I}{\partial \Phi} 
  + \frac{\Phi^* I}{M^2}\right|_{\Phi=\Phi_0} = 0\ ,
\end{equation}
and setting the present cosmological constant to be zero,
\begin{equation}
 V_{I} (\Phi_0) = 0\ ,
\label{Lambda}
\end{equation} 
implies
\begin{equation}
 I (\Phi_0) = \frac{\partial I}{\partial \Phi} (\Phi_0) = 0\ .
\end{equation}
Thus $I$ can be expanded as a Taylor series about its minimum.  The
{\em simplest} form for $I$ which satifies the above conditions
is~\cite{hrr84}
\begin{equation}
 I = \frac{\Delta^2}{M} (\Phi - \Phi_0)^2\ ,
\label{sic}
\end{equation}
where $\Delta$ is a mass parameter setting the energy scale for
inflation. Now in order for successful inflation to occur by the slow
roll-over mechanism, the scalar potential must be flat at the origin,
\begin{equation}
 \frac{\partial V_{I}}{\partial \Phi} |_{\Phi = 0} = 0\ ,
\end{equation}
which sets $\Phi_0 = M$. This in turn sets
\begin{equation}
 \frac{\partial^2 V_{I}}{\partial \Phi^2} |_{\Phi = 0} = 0\ ,
\end{equation}
since $I$ does not contain cubic terms. (The flatness is a
gauge-invariant property of the potential as $\Phi$ is a
gauge-singlet; indeed $\Phi$ cannot carry any quantum numbers because
it appears linearly in $I$.) The scalar potential obtained from
eq.~(\ref{VI}) is shown in Figure~1. The complex direction is stable
while along the real direction we can expand
\begin{equation}
 V_{I} (\phi) = \Delta^4 \left[1 - 4 \left(\frac{\phi}{M}\right)^3 
                 + \frac{13}{2} \left(\frac{\phi}{M}\right)^4  
                 - 8 \left(\frac{\phi}{M}\right)^5
                 + \frac{23}{3} \left(\frac{\phi}{M}\right)^6 
                 + \ldots \right]\ .           
\label{ourpot}
\end{equation} 
Can such a form arise naturally? We noted earlier (see eq.\ref{I})
that it was natural for the superpotential to acquire an overall mass
scale of this form due to the underlying symmetries of the theory.  We
also argued that the initial value of the field, $\phi_0$, corresponding
to the point at which the first derivative vanished would dominate the
final state of the universe. Expanding $V(\Phi/M)$ about this value,
\begin{equation}
 V(\Phi/M) = a\,(m) + b\,(m)(\Phi-\Phi_i) + c\,(m)(\Phi-\Phi_i)^2+ \ldots\ ,
\label{f}
\end{equation}
where the coefficients depend on the moduli $m$. The coefficient
$a(m)$ determines the value of the potential initially and hence the
moduli will flow to minimize this. However if the other coefficients
depend on independent combinations of the moduli they will be
undetermined at this stage as they do not affect the initial vacuum
energy. It is usually implicitly assumed that this is {\em not} the
case, since all the moduli are supposed to get SUSY-breaking masses
from gravitational coupling to the vacuum energy during
inflation. However we stress that this is not likely for the random
initial conditions assumed here. For example the simple K\"ahler
potential $K=\sum_j\Phi_j^{\dagger}\Phi_j$ leads to the potential
\begin{equation} 
 V_{I} = \e^{\sum_j|\Phi_j|^2/M^2}
  \left[\sum_k\left|\frac{\partial I}{\partial \phi_k} 
  + \frac{\phi_k^*I}{M^2}\right|^2 - \frac{3 |I|^2}{M^2}\right] .
\label{vj}
\end{equation}
During inflation one term in the square brackets must be non-zero and
thus one expects all fields $\phi_i$, including moduli fields, to get
a positive mass squared term. Minimizing this potential will fix all
the moduli vevs independently during inflation. However this form, if
it applies at all, will do so only about an enhanced symmetry point
corresponding to vanishing vevs for the $\Phi_i$. For initial
conditions far from this point, $K$ (and the corresponding exponential
in eq.\ref{vj}) will be a function of
$\Phi_j\langle\Phi_j^*\rangle+\Phi_j^*\langle\Phi_j\rangle$ and
$\Phi_j^{\dagger}\Phi_j$, so minimization will lead to a correlated
relation between the moduli fields, leaving other independent
combinations of moduli unconstrained. As discussed earlier, we then
expect that the initial conditions will allow {\em some} region in
which the value of $b\,(m)$ is just that needed to make the second
derivative of the potential vanish and this will dominate the final
state of the universe because of the enhanced amount of inflation it
will undergo. Thus we see that most of the features incorporated in
eq.(\ref{ourpot}) following from eq.(\ref{sic}) are indeed natural and
to be expected in any theory which has a potential with a turning
point. The exceptional property is that $I$ is a perfect
square. Recall that this was required to ensure the vanishing of the
cosmological constant at the minimum (see eq.\ref{Lambda}). By
adjusting the term $c\,(m)$ in eq.(\ref{f}) we can always arrange that
$V$ vanishes, but not in a natural way.

However it is not difficult to construct an alternative model which
does not suffer from this problem at all. All that is needed is a
reason why, at the end of inflation, the superpotential is at least
quadratic in the inflaton field expanded about the minimum of the
potential. This suggests an underlying symmetry, $Z_2$ or larger. The
example above does not have this symmetry because the K\"ahler
potential is not quadratic about $\Phi=M$. It is straightforward to
construct a variant which does have such a $Z_2$ symmetry via the
superpotential $P = \Delta^2 M (\Phi^2 + a\Phi^4 + \O(\Phi^6) \ldots)$
and a K\"ahler potential $K=\Phi^{\dagger}\Phi$. Now clearly the
associated potential vanishes at the minimum as required. In this case
the coefficient $a$ must be fine-tuned to give a potential with a
turning point away from $\Phi=0$ (which will now be the initial value)
at which the second derivative also vanishes. However, as stressed
above, in the case that $a$ is determined by moduli fields, the fine
tuned value will {\em automatically} be selected when considering the
most likely state of the universe. We will not pursue this example
further since the model already introduced contains the significant
features of this type of inflationary model. Thus henceforth we will
use the model of eqs.(\ref{sic}) and (\ref{ourpot}), which has this
cancellation, as we proceed to consider the reheat phase.

Integrating eq.(\ref{evolve}) back from the end of inflation, the
field value corresponding to the spatial scales probed by COBE is
$\phi_{1}\simeq~M/[12(N_1+2)]$ using $V'/V\simeq-12\phi^2/M^3$ from
eq.(\ref{ourpot}), so
\begin{equation}
 \epsilon_{1} \simeq 72 \left(\frac{\phi_1}{M}\right)^4, 
\end{equation}
which is $4.4\times10^{-10}$ for $N_1=51$, i.e. as small as is
required to solve the gravitino problem. (Note that for the ubiquitous
$\lambda\phi^4$ potential in chaotic inflationary scenarios, such a 
small slope can only be obtained for $\phi\geqsim~10^5~M$!)
The total number of e-folds of inflation is
\begin{equation}
 N_{\tot} \equiv \ln \frac{a(\phi_{\en})}{a(\phi_{\in})} 
   = \frac{1}{M^2} \int_{\phi_{\in}}^{\phi_{\en}} -\frac{V}{V'} \df\phi
   \simeq \frac{M}{12\Delta}\ ,
\end{equation}
setting $\phi_{\in}\sim\Delta$. Thus we need
$\Delta/M\leqsim1.7\times10^{-3}$ to get $N_{\tot}$ as large as is
neccessary to solve the cosmological horizon/flatness problems.  Since
the potential~(\ref{ourpot}) is dominated by a {\em cubic} term, the
spectrum of scalar density perturbations will depart from
scale-invariance at large $k$ as
$\sim\ln^2(Hk^{-1})_{\star}$~\cite{enot83}. This corresponds to a
`tilted' spectrum with slope given by
$n=1+2\eta_{1}-6\epsilon_{1}$~\cite{ll94}, so for the
potential~(\ref{ourpot}), we get
\begin{equation}
 n \simeq \frac{N_{1}-2}{N_{1}+2} \simeq 0.9\ ,
\end{equation}
which has less power on galactic scales in a CDM cosmogony and thus
will provide a better match to
observations~\cite{ssw95}.\footnote{Note that the tilt of the spectrum
is {\em not} associated with a significant tensor component in the CMB
anisotropy (although the converse is true~\cite{grav}), as is also the
case with the `natural' inflation model~\cite{natural}.} Because of
the tilt, the normalization to the COBE anisotropy is now somewhat
higher corresponding to
$\delta_{\H}\simeq2.5\times10^{-5}$~\cite{bsw95} for the perturbation
amplitude at large scales. This gives
\begin{equation}
 \frac{\Delta}{M} \simeq 1.4\times10^{-4},
\label{deltasic}
\end{equation} 
i.e. the inflationary scale is about $3\times10^{14}\gev$ and the
inflaton mass is about $5\times10^{10}\gev$. The reheat temperature is
then
\begin{equation}
 T_{\R} \sim 1.5\times10^{5} \gev,
\end{equation}
well below the conservative bound~(\ref{conTR}) and consistent with
even the recently proposed bound~(\ref{newTR}).

\subsection{Direct gravitino production\label{direct}}

It is not sufficient to inhibit the thermal production of gravitinos;
it is necessary, in addition, to suppress their production via
inflaton decay. This is a particularly important for an inflaton in
the hidden sector for all its couplings are gravitational and
therefore comparable to all states. Thus the solution to the reheating
constraint proves to be the problem where direct production is
concerned. The couplings to the gravitino can indeed be suppressed but
this is very model dependent. The best we can do here is to illustrate
the problem and present a possible solution in the context of the
model introduced above.

The couplings of the inflaton with interactions specified by the
superpotential of eq.(\ref{sic}) to the gravitino is contained in the
general coupling
\begin{equation}
 \frac{|I_s|}{M^2} \bar\psi_{3/2}^{\mu} \sigma_{\mu \nu}\psi_{3/2}^{\nu}\ ,
\end{equation}
where $I_s$ is the superpotential with superfields replaced by their
scalar components. This leads to the coupling
\begin{equation}
 h_{\phi \psi_{3/2} \psi_{3/2}} = 2\frac{\Delta^2 (\phi-M)}{M^3}\ .
\label{hgrav} 
\end{equation}
On the other hand the coupling of the inflaton to the scalar
components of matter fields may be read from eq.(\ref{ourpot}). The
dominant coupling comes from the interference term when the first term
is expanded, giving
\begin{equation}
 (\phi^* P_s + \phi P^*_s) \frac{2\Delta^2}{M^2}\ ,
\end{equation}
which contains quartic scalar couplings involving the inflaton and the
matter fields in the full superpotential. In the MSSM the dominant
coupling will be to the top squarks and the Higgs,
\begin{equation}
 h_{\phi^* \tilde{t} \tilde{t^c} H_2} = h_t \frac{2\Delta^2}{M^2}\ ,
\end{equation}
where $h_{t}$ is the top Yukawa coupling. We may use these couplings
to deduce the relative production of gravitinos in inflaton decay. The
crucial point is that although both couplings are gravitational in
origin, there is a suppression factor $(\phi-M)$ in the gravitino
coupling which follows because of the {\em square} appearing in the
superpotential $I$. As discussed earlier, this is dictated by the
empirical requirement that there be no cosmological constant after
inflation and may be expected in any model in which the contribution
to the cosmological constant from the inflationary sector
vanishes. The inflaton decays at a time given by eq.(\ref{tR}), so
using the virial theorem we may determine the mean value of the factor
in eq.(\ref{hgrav}) to be
\begin{equation}
 \langle(\phi-M)^2\rangle = \frac{\Delta^4}{16 \pi^3 M^2}\ ,
\end{equation}
giving
\begin{equation}
\frac{\Gamma_{\psi_{3/2} \psi_{3/2}}}{\Gamma_{\tilde{t} \tilde{t^c} H_2}}
 \sim \frac{\Delta^4}{M^4}\ .
\end{equation}
Is this suppression sufficient? The gravitinos produced by inflaton
decay are highly relativistic since $m_{\phi}\sim5\times10^{10}\gev$.
Until they become non-relativistic, their energy density relative to
that of radiation remains constant. However after they do, and until
they decay (or the universe becomes matter dominated) their relative
energy density grows linearly with the expansion factor. This amounts
to a growth of approximately $10^{9}$ for a TeV mass gravitino. Thus
the condition $\Delta/M\leqsim6\times10^{-3}$ is adequate to make
direct gravitino production phenomenologically acceptable. We see that
this is amply satisfied by the value in eq.(\ref{deltasic}) fixed by
the normalization to COBE.

\subsection{The Polonyi problem}

Lastly we consider whether a decoupled field will have stored
potential energy during inflation which is released very late, thus
recreating the problems associated with such a field that inflation
was supposed to cure~\cite{polprob,inflmass}. The fields of concern
abound in compactified (string) models~\cite{moduliprob}; they include
the moduli mentioned above which are responsible for determining the
couplings in the theory, the field related to the dilaton which
determines the gauge coupling, and the moduli fixing the K\"{a}hler
structure determining the shape and radius of compactification. The
precise details of this problem are model dependent but we shall
present a general description of the problem and determine under what
conditions one may hope to avoid it. A detailed exposition of this
question has recently been given in ref.~\cite{modulicosm}. While we
are broadly in agreement, our emphasis on the likely solution is
different and we have a new proposal as to how to deal with the moduli
(as distinct from the dilaton). The question of weak scale moduli has
also been recently discussed
elsewhere~\cite{dvali,2scale,ls95,thomas}; here we include a
discussion of much lighter moduli of the type which arise if they are
to determine the couplings of the Standard Model through their effect
on the electroweak effective potential~\cite{dynam}.

Suppose, as is expected in realistic models, that there is a stage of
spontaneous symmetry breaking below the Planck scale and denote by
$\delta$ the scale of the relevant potential. Now the moduli, $m$,
which determine the couplings of the theory enter this potential only
in the form of higher dimension operators suppressed by inverse powers
of the Planck mass:
\begin{equation}
 V (m) = \delta^4\ k \left(\frac{m}{M}\right)\ .
\label{Vm}
\end{equation}
where $k$ is some function. If $\delta$ is the largest scale of
spontaneous symmetry breaking then the vevs of the moduli are obtained
by minimizing this potential. Of course it may be that this is not
sufficient to determine all moduli, in which case one must consider
further sectors of spontaneous breaking, as in models~\cite{dynam}
where the couplings of the Standard Model are determined in this
manner by the electroweak symmetry breaking potential.

The properties of the moduli field are very similar to those of the
inflaton discussed above; its couplings are of gravitational strength
and its mass is very small, suppressed by inverse powers of the Planck
mass. Thus the moduli will have a lifetime given by
\begin{equation}
 \Gamma_m \sim \frac{(m_m)^3}{M^2} \sim \frac{\delta^6}{M^5}\ .
\end{equation}
We are concerned with the case $\delta<\Delta$ for then the moduli
will decay, after inflation, at very late times and release too much
entropy if the energy stored in the field is large. To estimate the
magnitude of this stored energy we note that the moduli typically
acquire a mass of $\O(\Delta^2/M)$ during
inflation~\cite{inflmass}.\footnote{It is possible to suppress this at
tree level in specific supergravity models~\cite{stewart} but
typically the canonical order of magnitude will be restored in
radiative order.} The sign of this term depends on the theory; if
one takes the canonical kinetic term it will be positive and tend to
stabilise the potential around $m=0$ but if negative it will drive a
large vev~\cite{br87}. The expectations for the two cases are,
respectively:
\begin{eqnarray}
 \langle{m}\rangle_+ \sim M \left(\frac{\delta^4}{\Delta^4}\right)
  k'(0)\ , 
  \nonumber\\ 
 \langle{m}\rangle_- \sim M \left(\frac{\Delta^4}{\delta^4}\right) 
  k''''(0)\ .
\label{vevc}
\end{eqnarray}
In the second case, minimization of the simple form for the potential
assumed above requires the moduli to be driven to large values, in
excess of the Planck scale. This is probably unrealistic and all we
will assume in this case is that the moduli are large, of $\O(M)$, but
dependent on $\Delta$. After inflation the vev of the moduli will
clearly not depend on $\Delta$. There are really only two
possibilities, viz. $\langle{m}\rangle=0$ or
$\langle{m}\rangle\sim\O(M)$. The former can be shown to be a local
minimum, for the moduli carry quantum numbers under discrete and/or
other symmetries and the vanishing vev corresponds to a point of
enhanced symmetry~\cite{ross}.

We can now quantify the problem. The energy stored in the potential
during inflation and released afterwards as the moduli flow to their
minima will be of $\O(\delta^4)$ {\em unless} the positions of the
minima during and after inflation coincide to an accuracy much better
than $M$. For the second possibility of eq.(\ref{vevc}), it is
unreasonable to suppose that the minima coincide to this accuracy. For
the first possibility, the minima could coincide but {\em only} if
after inflation the moduli vev remain zero corresponding to the
enhanced symmetry point. Apart from this possibility, the energy in
the moduli relative to the inflaton, just as inflation ends, is
$R_m\sim(\delta/\Delta)^4$. However the important point to note is
that the moduli are {\em not} in thermal equilibrium and will evolve
according to their equation of motion analogous to eq.(\ref{evolve}).
From this it follows that the evolution of the field towards its
minimum starts only at $t_{\en}$ when $H\sim\delta^2/M$. Until then,
the energy in the moduli field is in the form of potential energy and
thus grows relative to the energy in the inflaton field (or the
products of the inflaton after reheating). At $t_{\en}$, the moduli
roll rapidly to their minimum, converting the potential energy into
kinetic energy. However at this point $R_m\sim~1$, following
immediately from the condition that the roll starts. Now both the
inflaton and moduli are non-relativistic, so $R_m$ remains constant
until the time given in eq.(\ref{tR}) when the inflaton decays.
Thereafter until the moduli decay (or the (non-moduli) universe
becomes matter dominated), there will be a relative growth in the
moduli energy proportional to the scale-factor. Following
ref.\cite{moduliprob} this can be estimated to be,
\begin{equation}
 R_m = \min\left[\frac{\Delta^4}{\delta^4},
      \left(\frac{\Delta^6}{10^{-56} M^4}\right)^{2/3}\right]\ .
\label{eg}
\end{equation}
Clearly, any such growth is unacceptable as the universe will be
matter dominated by the moduli, hence the entropy released by their
subsequent decay will be far too large. We can envisage just three
possible solutions to this problem:

\begin{description}

\item(i) The moduli are fixed by a stage of symmetry breaking {\em
    before} inflation, i.e. $\delta>\Delta$. For example, new
    non-perturbative effects at a scale above that responsible for
    SUSY breaking may generate large moduli masses; however in all
    known examples these correspond to theories with negative
    cosmological constant, hence the universe necessarily suffers
    rapid gravitational collapse and does not evolve to a large enough
    size~\cite{modulicosm}. This does not exclude the idea altogether
    because there may well be string theories with such large
    non-perturbative effects and vanishing cosmological constant, but
    certainly one should find an example before concluding that this
    possibility is viable. Here we would like to point out a simple
    mechanism for generating moduli masses much larger than the SUSY
    breaking scale. We know that a large scale of symmetry breaking
    may be triggered by SUSY breaking along $D$- and $F$- flat
    directions. This comes about because a SUSY breaking scalar
    mass-squared for a field $\rho$ which may be positive at the
    Planck scale can be driven negative at a lower scale, $M_X$, by
    radiative corrections. This will trigger symmetry breaking along a
    $D$- and $F$- flat direction (if there is one). The important
    point is that the vev induced is energetically favoured to lie
    close to $M_X$, {\em independent} of the SUSY breaking scale, even
    though it acts as the trigger. This vev in turn can generate
    moduli masses of $\O(M_X^2/M)$ from a term in the superpotential
    of the form $(m m'/M^2)\rho^3$, where the quadratic structure
    applies near the point of enhanced symmetry~\cite{ross}. Thus we
    consider it quite likely that theories with a large intermediate
    scale of symmetry breaking will have moduli masses sufficiently
    large to avoid the Polonyi problem. However this mechanism does
    not work for the dilaton whose couplings have a different
    character. Although it is unlikely that the dilaton should acquire
    a mass much higher than the electroweak scale~\cite{modulicosm},
    it is not proven that this cannot happen through non-perturbative
    effects; indeed it has been conjectured that this does indeed
    happen~\cite{lnl95}. In our opinion this would be the best
    solution, since otherwise it appears impossible to construct an
    inflationary potential at all. This is because in superstring
    theories the potential necessarily has a dependence on the dilaton
    and, with a light dilaton, the curvature of the potential in the
    dilaton direction is too large to permit
    inflation~\cite{stringinfl}.

\item(ii) The moduli minima are the {\em same} during and after
  inflation. As stressed above, this is quite reasonable but only if
  this corresponds to a point of enhanced symmetry. The example given
  above illustrates how this comes about. However this explanation
  will not allow scenarios in which the moduli and associated
  Yukawa or soft SUSY breaking masses are determined at a low
  scale such as the electroweak breaking scale~\cite{dynam}, because in
  this case the moduli vevs adjust in response to low energy phenomena
  (such as the alignment of soft masses with the fermion masses) which
  arise only on electroweak breaking. It seems exceedingly unlikely
  that the low energy minimum would, in this case, correspond to the
  minimum of the moduli during inflation.

\item(iii) The cosmology is non-standard after $t_{\R}$. It has been
  suggested that there may have been a brief secondary period of
  inflation (before baryogenesis) during which the moduli roll to
  their minima~\cite{2scale}; the following reheat phase then dilutes
  the moduli energy. (This is effectively equivalent to the
  possibility $\delta>\Delta$ but with a smaller $\Delta$.) To be
  viable, this scenario must be carefully constructed --- the second
  epoch of inflation must be short enough not to erase the density
  perturbations produced in the initial inflationary era, but long
  enough to adequately dilute the moduli energy. Now if the moduli are
  to roll to their minimum, the Hubble parameter during the second
  phase of inflation must be less than the moduli mass. For the case
  of moduli with electroweak scale masses considered in
  refs.\cite{2scale,ls95} this requires the vacuum energy during
  inflation to be at an intermediate scale of
  $M_{\I}^4\sim\,m_{W}^2M^2$. There should be fewer than $\sim25-30$
  e-folds of inflation if the density perturbations generated in the
  first stage of inflation are not to be erased. Finally the reheat
  phase must proceed through unsuppressed renormalizable couplings to
  Standard Model states in order to avoid reintroducing the moduli
  problem via the new inflaton. A particularly promising scenario for
  this second stage of inflation has been suggested
  recently~\cite{ls95}. It is noted that thermal effects can keep the
  field value at the origin from the temperature,
  $T_1\sim(m_{W}M_{\I})^{1/2}$, at which the potential dominates,
  until the temperature, $T_2\sim\,m_{W}$, at which the supersymmetry
  breaking mass triggers the phase transition. Thus the number of
  e-folds of inflation is automatically limited. Further it is
  plausibly argued that the relatively low scale of symmetry breaking
  needed to solve the moduli problem (i.e. $M_{\I}\sim10^9\gev$) is
  expected in supersymmetric theories with flat directions. Such an
  epoch of `thermal' inflation is expected to occur irrespective of
  previous stages of inflation. Thus there is no conflict with our
  criterion of naturalness, i.e. we do not require multiple {\em
  correlated} periods of inflation. However, while this may cure the
  problem for moduli with weak scale masses, it cannot do so for much
  lighter moduli because the energy stored in such fields is not
  released until the Hubble parameter falls below their mass and they
  start to roll. In particular the moduli whose vevs are determined at
  the electroweak scale have masses of $\O(m_{W}^2/M)$, and their
  energy will {\em not} be diluted by inflation at an intermediate
  scale. To invoke the solution of refs.\cite{2scale,ls95} one now
  needs inflation at the weak scale with an inflationary potential of
  $\O(m_{W}^4)$. However the reheat temperature is then far below the
  value of $\sim10\mev$ needed to avoid disrupting standard
  nucleosynthesis. One can argue that along a flat direction the
  potential may be of $\O(m_{W}^4)$ while still having renormalizable
  couplings for the inflaton; however, following the argument
  presented in ref.~\cite{2scale}, the reheat temperature will still
  be rather low, $T_{\R}\sim~g^{2/3}m_{\phi}^{5/6}M^{1/6}$, where $g$
  is a gauge or Yukawa coupling. Taking $m_{\phi}\approx~m_{W}^2/M$,
  this yields a maximum reheat temperature of $\O(10^{-6})\mev$! Given
  this problem we know of no way to reconcile moduli of mass
  $\O(m_W^2/M)$ with the cosmological constraints discussed above if
  their vevs are determined at the electroweak breaking scale.

\end{description}

\section{Conclusions}

We have re-examined the possibilities for and the problems associated
with supersymmetric inflation of the type that one might expect in
compactified string theories. In the effective supergravity theory
following from such theories below the Planck scale, there are many
hidden sectors corresponding to fields with only gravitational
strength couplings. Such sectors offer very plausible inflaton
candidates because the fields have rather small masses, related to
spontaneous symmetry breaking below the compactification scale. As a
result, inflation, if it occurs, will be associated with a scale far
below the Planck scale, just as is required if an acceptable spectrum
of density perturbations is to be generated. Whether there is an
inflationary era at all in such theories depends on the initial
conditions and the form of the effective potential. We have argued
that weakly coupled fields which are not in thermal equilibrium are
likely to have random initial conditions. As a result the final state
of the universe is dominated by regions which have undergone an
inflationary era, corresponding to a choice of initial conditions of
the moduli for which the potential is anomalously flat and to an
initial value of the inflaton near the flat part of its potential.
Thus we consider that all the ingredients of an acceptable
inflationary potential are present in effective supergravity theories
without the need to fine tune the parameters or to seek forms of the
potential such as in the chaotic inflation scenario which inflate for
a wide range of initial conditions. An analysis of the possible origin
of the scale of inflation suggests that it should be bounded above by
the scale of SUSY breaking. If this arises through gaugino
condensation then the relevant scale may be of $\O(10^{13})\gev$.
Inflationary models with a flat potential and sub-Planck scale vevs
require such an intermediate scale to generate the correct level of
density perturbations. On the other hand chaotic inflationary models
require a much larger scale.

The main difficulty encountered by supergravity inflationary schemes
concerns the cosmology {\em after} inflation. This is because the many
hidden sector fields which were desirable to give candidate inflatons
now cause severe problems. Due to their weak coupling they decay very
late in the evolution of the universe and, if they contain a sizeable
fraction of the total energy density, produce unacceptable amounts of
entropy in the process. We discussed three aspects of this problem in
some detail. The first is the thermal production of such weakly
coupled states, e.g. the gravitino, due to reheating after inflation.
We noted that the hidden sector inflaton avoids this problem in a very
natural way for, having only gravitational strength couplings, it
reheats the universe to a low temperature. In simple models with only
a single stage of inflation it is possible to relate the reheat
temperature directly to the spectrum of density perturbations.  We
found that normalizing the latter to the COBE data constrains the
reheat temperature to be $\sim10^{5}\gev$, comfortably within the
phenomenological bounds for a TeV scale gravitino. An important, and
observationally testable, implication is that there should be a
negligible tensor component in the CMB anisotropy and that the power
spectrum of scalar density perturbations should be tilted away from
scale-invariance. The models which provide this nice consistency have
small values of the inflaton field relative to the Planck scale. In
contrast, models of the chaotic type (e.g. ref.\cite{susychaotic})
give too high a reheat temperature if they are normalized to generate
the correct amplitude of density perturbations. Of course it is
possible to evade these conclusions by postulating non-standard
evolution after the inflationary era but such schemes may require a
measure of fine tuning if they are not to destroy the density
perturbations produced by the initial stage of inflation.\footnote{As
just discussed, an exception is thermal
inflation~\cite{ls95}.} Moreover, given the success of the simplest
schemes, it seems to us reasonable to consider their predictions as
the paradigm, rather than those of more involved models,
e.g. involving two coupled fields~\cite{2fields}, which are
constructed to avoid problems not encountered in the simple ones.

The second potential problem for the supergravity inflation models is
the direct production of gravitinos and other weakly coupled states
through inflaton decay. This is more model dependent than thermal
production and we have considered this problem only for minimal
supergravity. We find that the {\em empirical} requirement that there
be vanishing cosmological constant in the inflaton sector after
inflation provides the required suppression of the production of
gravitinos relative to matter fields. Again we find that the bounds
for a TeV mass gravitino are easily satisfied for a low inflationary
scale.

Finally we considered the Polonyi problem, viz. the difficulty in
suppressing the potential energy stored during inflation in weakly
coupled hidden sector fields such as the moduli of compactified string
theories. We considered three possible solutions. The first is that
all moduli have vevs fixed at a scale above inflation. The second is
that the minimum of the potential during inflation coincides with the
minimum after inflation. The third is a late stage of inflation. In
all cases the implication is that the moduli {\em cannot} be treated
as dynamical variables at the electroweak scale determining the
couplings in the low energy theory.

The low reheat temperature requires the baryon asymmetry of the
universe to be generated at a relatively low energy scale and there is
a natural mechanism~\cite{susybaryogen} to achieve this in the context
of supergravity inflation. We are presently examining this question as
well as the detailed expectations in our model for observations of
large-scale structure~\cite{lss}.

\vspace{5mm}
\noindent
{\bf \Large Acknowledgement:} We wish to thank Steven Abel,
Maria Bento, Orfeu Bertolami, Pierre Bin\'{e}truy, Savas Dimopoulos,
Gian Giudice, Costas Kounnas, Fernando Quevedo, Lisa Randall, Ewan
Stewart and Fabio Zwirner for helpful comments and criticism. We are
particularly grateful to Andrei Linde and David Lyth for clarifying
discussions.

\newpage
\begin{figure}[h]
\vskip 9in\hskip-1in\includegraphics{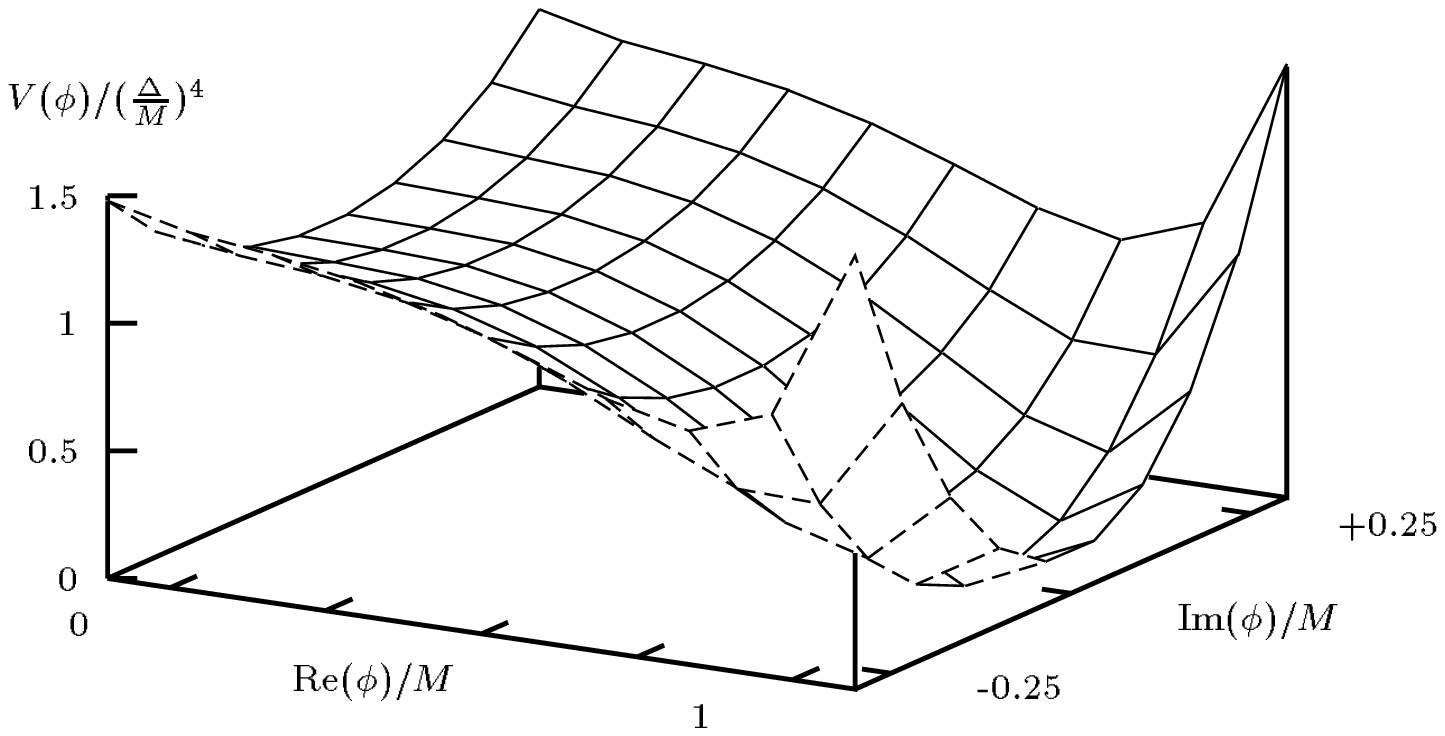}
\vskip-3in\caption{The complex scalar inflationary potential.}
\end{figure} 

\end{document}